\documentclass[manuscript]{aastex}
\usepackage{natbib}
\newcommand{\fexxi}{Fe \scriptsize{XXI} \normalsize}
\newcommand{\fexii}{Fe \scriptsize{XII} \normalsize}
\newcommand{\feii}{Fe \scriptsize{II} \normalsize}
\newcommand{\ci}{C \scriptsize{I} \normalsize}
\newcommand{\siiv}{Si \scriptsize{IV} \normalsize}
\newcommand{\siii}{Si \scriptsize{II} \normalsize}
\newcommand{\oiv}{O \scriptsize{IV} \normalsize}
\newcommand{\oi}{O \scriptsize{I} \normalsize}
\newcommand{\si}{S \scriptsize{I} \normalsize}
\newcommand{\sxv}{S \scriptsize{XV} \normalsize}
\newcommand{\caxix}{Ca \scriptsize{XIX} \normalsize}

\begin{document}

\title{Imaging and Spectral Observations of Quasi-Periodic Pulsations in a Solar Flare}

\author{D. Li\altaffilmark{1,2}, Z. J. Ning\altaffilmark{1},
        and Q. M. Zhang\altaffilmark{1}}
\affil{$^1$Key Laboratory of Dark Matter and Space Astronomy, Purple
Mountain Observatory, CAS, Nanjing 210008, China}
\affil{$^2$University of Chinese Academy of Sciences, Beijing
100049, China}
\altaffiltext{3}{Correspondence should be sent to:
lidong@pmo.ac.cn.}

\begin{abstract}
We explore the Quasi-Periodic Pulsations (QPPs) in a solar flare
observed by {\it Fermi} Gamma-ray Burst Monitor (GBM), {\it Solar
Dynamics Observatory} ({\it SDO}), {\it Solar Terrestrial Relations
Observatory} ({\it STEREO}), and {\it Interface Region Imaging
Spectrograph} ({\it IRIS}) on 2014 September 10. QPPs are identified
as the regular and periodic peaks on the rapidly-varying components,
which are the light curves after removing the slowly-varying
components. The QPPs display only three peaks at the beginning on
the hard X-ray (HXR) emissions, but ten peaks on the chromospheric
and coronal line emissions, and more than seven peaks (each peak is
corresponding to a type III burst on the dynamic spectra) at the
radio emissions. An uniform quasi-period about 4 minutes are
detected among them. AIA imaging observations exhibit that the 4-min
QPPs originate from the flare ribbon, and tend to appear on the
ribbon front. {\it IRIS} spectral observations show that each peak
of the QPPs tends to a broad line width and a red Doppler velocity
at \ci, \oiv, \siiv, and \fexxi lines. Our findings indicate that
the QPPs are produced by the non-thermal electrons which are
accelerated by the induced quasi-periodic magnetic reconnections in
this flare.
\end{abstract}


\section{Introduction}
Quasi-Periodic Pulsations (QPPs) are regular phenomenons and common
features observed in the solar flare emissions. In a typical event,
QPP displays as the periodic peaks on the light curve. Each peak has
a similar lifetime, which results in a regular interval among them.
Therefore, the QPPs are characterized by the repetition or the
periodicity. They are observed with the typical periods ranging from
milliseconds \citep[e.g.,][]{Karlicky05,Tan10} through seconds
\citep[e.g.,][]{Hoyng76,Lipa78,Bogovalov83,Bogovalov84,Mangeney89,Zhao91,Aschwanden94,Ning05,Zimovets10,Nakariakov10}
to minutes
\citep[e.g.,][]{Nakariakov99,Aschwanden02,Demoortel02,Foullon05,Ofman06,Li08,Sych09,Tan10,Su12a,Ning14}.
The previous observations show that their periods are positively
correlated with the major radius of the flaring loop
\citep{Aschwanden98}. The short QPPs are thought to be associated
with kinetic processes caused by the dynamic interaction of
electromagnetic, plasma or waves with energetic particles trapped in
closed magnetic fields \citep{Aschwanden87,Nakariakov09}. The long
QPPs are usually associated with active region dynamics and global
oscillations of the Sun \citep{Chen06,Nakariakov09}.

The QPPs are detected in a broad wavelength range from radio
\citep[e.g.,][]{Mangeney89,Zhao91,Aschwanden94,Kliem00,Karlicky05,Ning05,Tan10}
through visible and extreme-ultraviolet (EUV)
\citep[e.g.,][]{Nakariakov99,Aschwanden02,Demoortel02,Ofman02,Su12a,Su12b}
to X-rays
\citep[e.g.,][]{Hoyng76,Lipa78,Bogovalov83,Bogovalov84,Foullon05,Ofman06,Nakariakov06,Li08,Zimovets10,Nakariakov10,Ning14},
and even to $\gamma$-rays \citep{Nakariakov10}. In radio emissions,
the QPPs are usually detected as the periodic type III bursts.
\citet{Mangeney89} have reported the periods range from 1 to 6 s.
The statistical studies of the quasi-periodicity in the normal and
reverse slope type III bursts have been reported by
\citet{Aschwanden94} and \citet{Ning05}, respectively. Both of them
obtained the mean period of $\sim$2 s and believed that the
periodicity is due to the periodic acceleration processes in the
solar flares. At EUV wavelengths, QPPs are usually detected in
coronal loops. Moreover, using the spectral observations, the QPPs
in Doppler velocity and line width of the hot and cool lines are
also detected \citep{Kliem02,Tian11}. For example, using Solar
Ultraviolet Measurement of Emitted Radiation (SUMER) spectrometer on
board {\it Solar and Heliospheric Observatory} ({\it SOHO}), the
Doppler velocity in hot lines ($>$ 6 MK) are detected to exhibit
QPPs with period of about ten minutes \citep{Ofman02,Wang02,Wang03}.
With the Bragg Crystal Spectrometer (BCS) on {\it Yohkoh}, the
Doppler velocity in the flare emission lines (e.g., \sxv and \caxix)
is observed to display the QPPs with the period of serval minutes
\citep{Mariska05,Mariska06}. \citet{Tian11} found that the QPPs are
correlated with line intensity, Doppler velocity, and line width
from the observations of {\it Hinode} EUV Imaging Spectrometer
(EIS). In the X-ray and $\gamma$-ray channels, the QPPs with a short
period of a few seconds are detected by \citet{Hoyng76} and
\citet{Bogovalov83}. Using {\it Reuven Ramaty High-Energy Solar
Spectroscopic Imager} ({\it RHESSI}) observations, the 2005 January
19 solar flare displays the QPPs with a period of 2$-$4 minutes at
HXR emissions \citep{Ofman06}, while the 2002 December 26 solar
flare exhibits the 2-min QPPs at the soft X-ray (SXR) emissions
\citep{Ning14}. Some flares exhibit similar QPPs in a broad
wavelengths \citep[e.g.,][]{Nakajima83,Aschwanden95,Asai01}. For
example, the 1998 November 10 solar flare displays the QPPs with a
period of 6 s in both radio and X-ray emissions \citep{Asai01},
while \citet{Nakajima83} reported the QPPs with a quasi-period of 8
s from radio, hard X-ray (HXR) and $\gamma$-ray emissions.

The generation mechanism of QPPs is still an open issue in the
documents
\citep{Aschwanden87,Nakariakov06,Li08,Nakariakov09,Ning14}.
Basically, QPPs are thought to be related with the waves or
energetic particles (electrons). Actually, MHD waves, such as slow
magnetoacoustic waves, fast kink and sausage waves \citep{Roberts84,
Nakariakov09} have been used to explain the generation of the QPPs
in radio \citep{Nakariakov04,Tan10,Kupriyanova10}, EUV
\citep{Ofman02,Mariska06,Tian11,Su12a,Su12b} and X-ray or $\gamma$-
ray \citep{Nakariakov04,Foullon05,Zimovets10,Nakariakov10}
emissions. On the other hand, the QPPs could be explained by the
emissions from the non-thermal electrons that are accelerated by the
quasi-periodic magnetic reconnection. And the quasi-periodic
reconnection can be spontaneous
\citep{Kliem00,Karlicky04,Karlicky05,Murray09} or may be modulated
by MHD waves, i.e., the slow waves
\citep{Chen06,Nakariakov11,Liting15} or the fast waves
\citep{Nakariakov06,Nakariakov09,Ofman06,Liu11}. Until now, QPPs are
poorly observed in a same flare with the imaging and spectral
observations simultaneously, which could provide an opportunity to
improve the QPPs origination and physics model. In this paper, we
analyze the QPPs in a solar flare on 2014 September 10 observed by
{\it Fermi}, {\it SDO}, {\it STEREO}, and {\it IRIS} at HXR, EUV,
and radio wavelengths.

\section{Observations and Data Analysis}
\begin{figure}
\epsscale{0.9}  \plotone{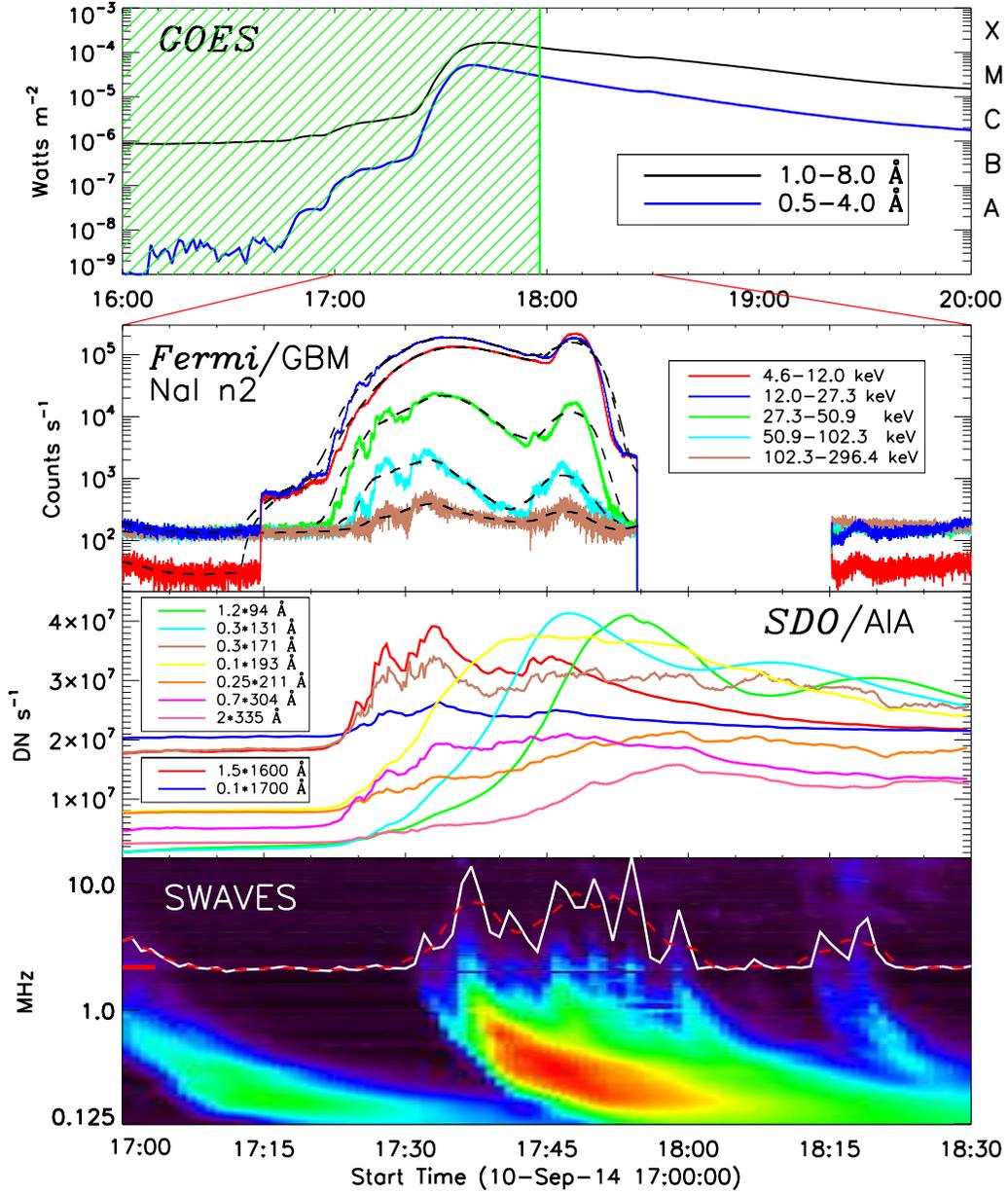} \caption{Top: {\it GOES} SXR flux
from 16:00 UT to 20:00 UT on 2014 September 10. The vertical green
line marks the end time of {\it IRIS} observations. Second: X-ray
light curves from {\it Fermi}/GBM (detector n2) at 5 energy
channels. The dashed lines represent the slowly-varying components.
Third: AIA light curves integrated from the flare region marked with
the blue box in Fig.~\ref{active}. Bottom: radio dynamic spectra
between 0.125$-$16.075 MHz from {\it STEREO}/WAVES. The white
profile is the radio emission at frequency of $\sim$2.19 MHz which
marked with the short red line, and the red dashed line represents
the slowly-varying components. \label{light}}
\end{figure}

The solar flare studied in this paper takes place in NOAA AR12158
(N11$^{\circ}$, E05$^{\circ}$) on 2014 September 10, and it is
accompanied with a halo coronal mass ejection (CME). It is an X1.6
flare, which starts at 17:21 UT and reaches its maximin at 17:45 UT
from {\it GOES} SXR flux. Fig.~\ref{light} shows the light curves
detected by {\it GOES}, {\it Fermi}/GBM, and {\it SDO}/AIA, the
dynamic spectra detected by {\it STEREO}/WAVES (SWAVES). The top
panel gives the {\it GOES} observations at two SXR channels, such as
1.0$-$8.0 {\AA} (black) and 0.5$-$4.0 {\AA} (blue). The shaded
interval marks {\it IRIS} observations. The second panel shows the
{\it Fermi} light curves at 5 energy channels, such as 4.6$-$12.0
keV, 12.0$-$27.3 keV, 27.3$-$50.9 keV, 50.9$-$102.3 keV, and
102.3$-$296.4 keV. They are detected by the n2 detector, whose
direction angle to the Sun is stable ($\sim$60$^{\circ}$) during
this flare, especially at the interval from 17:10 UT to 17:45 UT,
while the other detectors change their direction angles frequently.
After 17:45 UT, the n2 detector shifts its direction angle from
$\sim$60$^{\circ}$ to $\sim$45$^{\circ}$, then become bigger again,
which results into a X-ray peak around 17:50 UT. Therefore, it is
not the real X-ray emission peak. There is a data gap after 17:54
UT. The time resolution of {\it Fermi} is 0.256 s, but becomes to
0.064 s automatically in the flare state \citep{Meegan09}. We
interpolate all the data into an uniform resolution of 0.256 s in
the second panel. Such a cadence is enough to analyze the QPPs with
a period of several minutes in this paper. The third panel shows the
{\it SDO}/AIA light curves (integration from images) at 9
wavelengths, such as 1600 {\AA}, 1700 {\AA}, 94 {\AA}, 131 {\AA},
171 {\AA}, 193 {\AA}, 211 {\AA}, 304 {\AA}, 335 {\AA}. Their time
resolutions are 24 s here. The bottom panel displays the radio
dynamic spectra between $\sim$0.125 MHz and $\sim$16.075 MHz
observed by SWAVES aboard {\it STEREO}\_B. The time resolution is 1
minute \citep{Rucker05}. There is a group of solar radio type III
bursts.

Fig.~\ref{light} shows that there are several peaks during the
impulsive phase at HXR light curves, i.e., from 17:21 to 17:40 UT.
These peaks seem to be regular and periodic, and they looked like
the QPPs. However, they are superposed by a gradual background
emission. Meanwhile, there are several type III bursts on the
dynamic spectra. The radio light curve at 2.19 MHz (white line)
exhibits the regular peaks with a quasi-period. In order to
distinguish these peaks from the background, we decompose each light
curve at X-ray and radio bands into a slowly-varying component and a
rapidly-varying component. The slowly-varying component (the
background emission) is the smoothing original data. Here the
smoothing window is different for various data with the similar
cadences. For example, the smoothing window is 1000 points for {\it
Fermi} data and 4 points for SWAVES data. The dashed lines
over-plotted on the original light curves in Fig.~\ref{light} are
their slowly-varying components.

\begin{figure}
\epsscale{1.0} \plotone{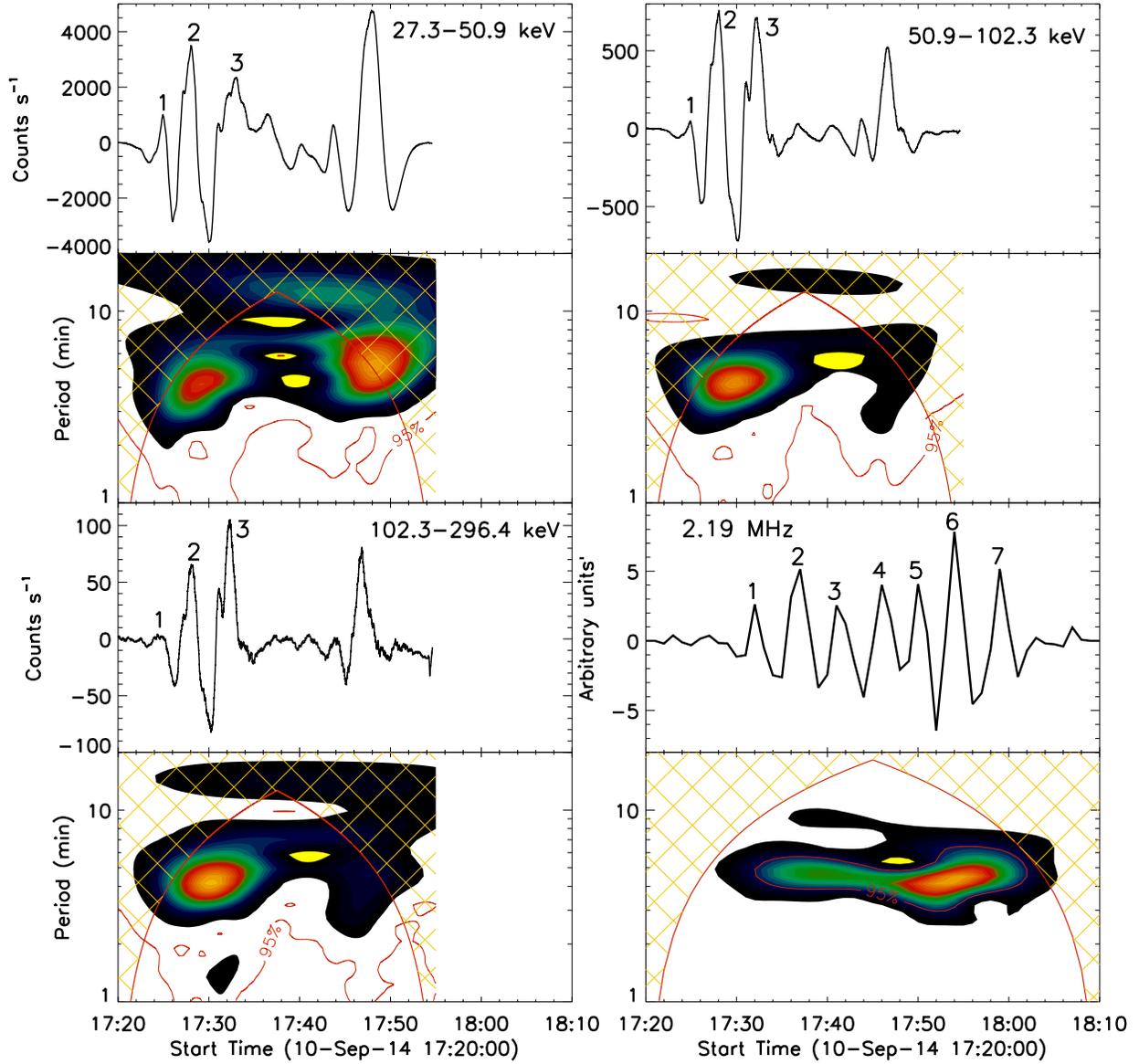} \caption{Top: the rapidly-varying
components at {\it Fermi} 27.3$-$50.9 keV, 50.9$-$102.3 keV,
102.3$-$296.4 keV, and radio emission at frequency of $\sim$2.19
MHz. Bottom: their wavelet power spectra. \label{qpp1}}
\end{figure}

QPPs are identified from the rapidly-varying components, which are
the light curves subtracted by the slowly-varying components. As
shown in Fig.~\ref{light}, we get the slowly-varying components at
five X-ray bands (black dashed lines) and at one radio frequency
(red dashed lines). Fig.~\ref{qpp1} gives the rapidly-varying
components at three HXR channels of 27.3$-$50.9 keV, 50.9$-$102.3
keV, and 102.3$-$296.4 keV and one radio frequency at $\sim$2.19
MHz. These three HXR channels display the typical QPPs with three
regular peaks which marked by the number of `1', `2', and `3'
between 17:24 UT and 17:36 UT. There are no similar peaks in the
other two X-ray bands below 27 keV. However, there are seven regular
peaks between 17:32 and 18:01 on the radio frequency at $\sim$2.19
MHz, marked by the numbers. Then the wavelet analysis is used to
detect the period of the QPPs. The bottom panels of Fig.~\ref{qpp1}
show their wavelet power spectra, which confirm the QPPs feature
with a similar period of about 4 minutes at both HXR and radio
emissions. As mentioned earlier, the HXR peaks at 17:50 UT are not
real, although they are the emissions from the Sun.

There are two interesting facts to be mentioned here. Firstly, the
HXR light curves display the QPPs from 17:24 UT to 17:36 UT, while
the radio emissions exhibit the QPPs from 17:32 UT to 18:01 UT.
Namely, there is a delay of about 8 minutes between their onset.
Secondly, the QPPs show three peaks at HXR channels, but more than
seven peaks at radio frequency. Each peak is a type III burst on the
dynamic spectra. The question is whether the QPPs at HXR are related
to that at the radio in this event, and whether the QPPs at HXR and
radio originate the same process during the flare? In order to
answer these questions, imaging and spectral analyses of the QPPs
are needed. High spatial and time resolutions images of {\it
SDO}/AIA and high spectral resolution of {\it IRIS} observations
give us the opportunity to study the QPPs origination in the 2014
September 10 flare.

Fig.~\ref{active} shows the {\it SDO}/AIA 1600 {\AA}, 131 {\AA} and
304 {\AA} images and the {\it IRIS}/SJ 1400 {\AA} images at 17:30 UT
before the flare maximum. Same as the other X-class flares, this
event displays double ribbons at 1600 {\AA}. One is short near the
sunspot, while the other one is long and shows a curved shape. The
light curves integrated in the blue box are given in the third panel
of Fig.~\ref{light} at all 9 AIA wavelengths. AIA has a pixel size
of 0$\farcs$6 and a time resolution of 24 s at 1600 {\AA} and 1700
{\AA}, while 12 s at the other 7 EUV wavelengths \citep{Lemen12}.
For the 2014 September 10 flare, AIA images at these seven EUV
wavelengths are regularly saturated every 24 s. These saturation
images are ruled out to do the analysis. The light curves at these 7
EUV wavelengths also have the time resolution of 24 s as same as
that at 1600 {\AA} and 1700 {\AA}. The slit-jaw images (SJI) aboard
{\it IRIS} take the solar images with a FOV (field of view) of
119$\arcsec$ $\times$ 119$\arcsec$ and pixel size of 0$\farcs$166.
The time cadence is 19 s at 1400 {\AA} and 2796 {\AA}. The right
upper panel in Fig.~\ref{active} gives one SJI at 1400 {\AA} and it
also includes the long ribbon of the solar flare. The AIA and SJ
images have been pre-processed with the standard solar-software
routines \citep{Marc13,Mclntosh13} to be aligned. The AIA images at
1600 {\AA} are used to co-align with the SJ images at 1400 {\AA}
because both of them include the continuum emissions from the
temperature minimum, and the continuum emissions are dominant in
many of the bright features. The upper middle and right panels in
Fig.~\ref{active} show the results of the co-alignment between AIA
1600 {\AA} and SJ 1400 {\AA} images at 17:30 UT. They have the same
scales, and similar bright features. The black line indicates the
{\it IRIS} slit position. The red box in the left panels marks the
FOV of the SJ image in the active region. Fig.~\ref{active} also
shows the same regions at AIA 131 {\AA} and AIA 304 {\AA}, which
correspond to the high and low temperatures, respectively. We make
the movie for Fig.~\ref{active} from 17:12 UT to 17:58 UT, which
contains the impulsive phase of the X1.6 flare (seen, f3.mpg).

\begin{figure}
\epsscale{1.0} \plotone{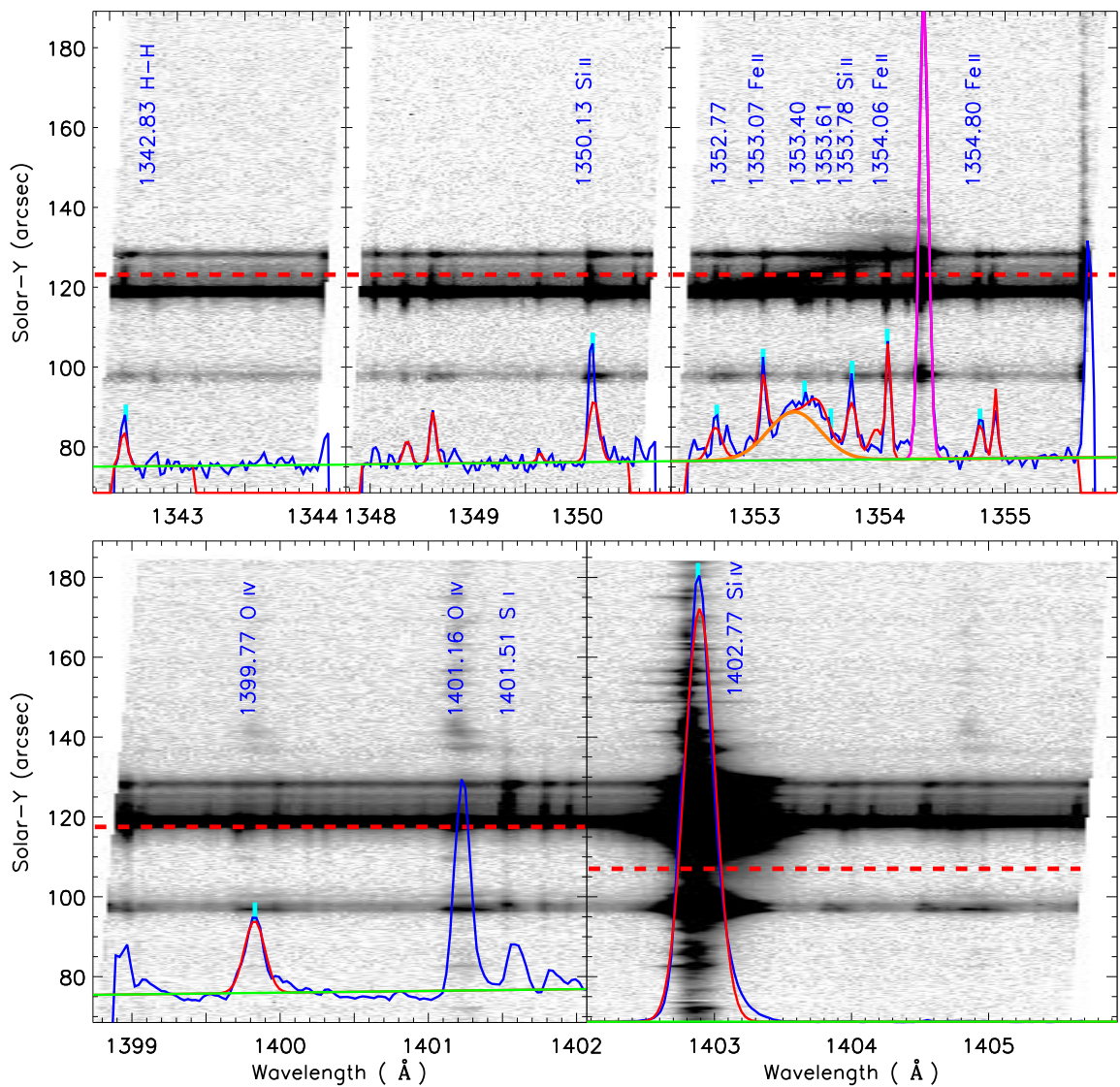} \caption{{\it IRIS} spectra at 17:30
UT from FUV1 (upper) (e.g., 1343, \fexii 1349 and \oi 1356) and FUV2
(bottom) bands (e.g., \siiv 1403). The blue spectral profiles are
from the flare ribbon at the slit positions marked by the red dashed
lines. The horizontal green lines mark the background profiles. The
orange profile is the \fexxi fitting, and the purple profile is the
\ci fitting. The main lines used to fit for this flare are labeled
and indicated by the turquoise vertical ticks just above the line
spectra. \label{fit}}
\end{figure}

{\it IRIS} spectrograph observes the AR12158 from 11:28 UT to 17:58
UT on 2014 September 10 in a ``sit-and-stare'' mode, and the step
cadence is $\sim$9.4 s. The pixel size along the slit is
$\sim$0$\farcs$166, and the spectral scale is $\sim$12.8
m{\AA}/pixel at FUV (FUV1 and FUV2) bands
\citep{Mclntosh13,Depontieu13,Depontieu14}. But to save telemetry,
two times spectral binning and a restricted number of spectral
windows were obtained, i.e., the spectral scale is 25.6
m{\AA}/pixel, equivalent to 5.6 km s$^{-1}$/pixel. In this case, we
used the `flare' list of lines which are consisted of the `1343,
\fexii, \oi, \siiv' windows. Fig.~\ref{fit} shows the {\it IRIS}
spectra of FUV1 (e.g., 1343, \fexii 1349 and \oi 1356) and FUV2
(e.g., \siiv 1403) windows at 17:30 UT. They have been processed to
remove the bad pixels. Four lines at \fexxi 1354.09 {\AA}, \ci
1354.33 {\AA}, \oiv 1399.77 {\AA}, and \siiv 1402.77 {\AA} are
selected to do analysis. The former two lines are in the FUV1
window, while the \oiv and \siiv lines are in the FUV2. It is well
known that the forbidden line of \fexxi 1354.09 {\AA} is a broad
line and it is always blended with other narrow lines from
chromospheric emissions, especially the chromospheric line of \ci
1354.33 {\AA}
\citep[e.g.,][]{Doschek75,Cheng79,Mason86,Innes03a,Innes03b}, which
makes it difficult to fit. {\it IRIS} has a high spectral resolution
of $\sim$26 m{\AA}, which results into distinguishing a lot of
bright emission lines. The upper panel of Fig.~\ref{fit} shows that
some of them are well identified, such as \ci 1354.33 {\AA} (purple
line), \feii 1353.07 {\AA} and 1354.06 {\AA}, \siii 1353.78 {\AA},
but some of them are still not identified, i.e., 1352.77 {\AA},
1353.40 {\AA}, and 1353.61 {\AA}. These known and unknown bright
emission lines which blend with the \fexxi line must be extracted
before determining the \fexxi intensity. In this case, the fit
method described by \citet{Li15} are used to extract the \fexxi line
information. Briefly, we fix these bright emission line positions,
constrained their widths and tied their intensities to the lines in
other spectral windows. Finally, 15 Gaussian lines superimposed on a
linear background are fitted across the whole wavelength region.
Only 11 lines are marked with the turquoise vertical ticks in the
upper panel of Fig.~\ref{fit}. Thus we can detect the integral
intensities, line widths and Doppler velocities of these lines
simultaneously in each fitting, including the chromospheric line of
\ci 1354.33 {\AA} and the coronal line of \fexxi1354.09 {\AA}. The
bottom panel of Fig.~\ref{fit} shows the FUV2 windows with four
lines, such as \oiv 1399.77 {\AA} and 1401.16 {\AA}, \si 1401.51
{\AA} and \siiv 1402.77 {\AA}. Two lines ( \oiv 1401.16 {\AA} and
\si 1401.51) are ruled out to analyze because they are blended with
each other, especially during the flare time. The other two lines
are isolated and can be well fitted with a single Gaussian function
(red lines) to detect the integral intensities, line widths and
Doppler velocities.

\section{Results}
\begin{figure}
\epsscale{1.0} \plotone{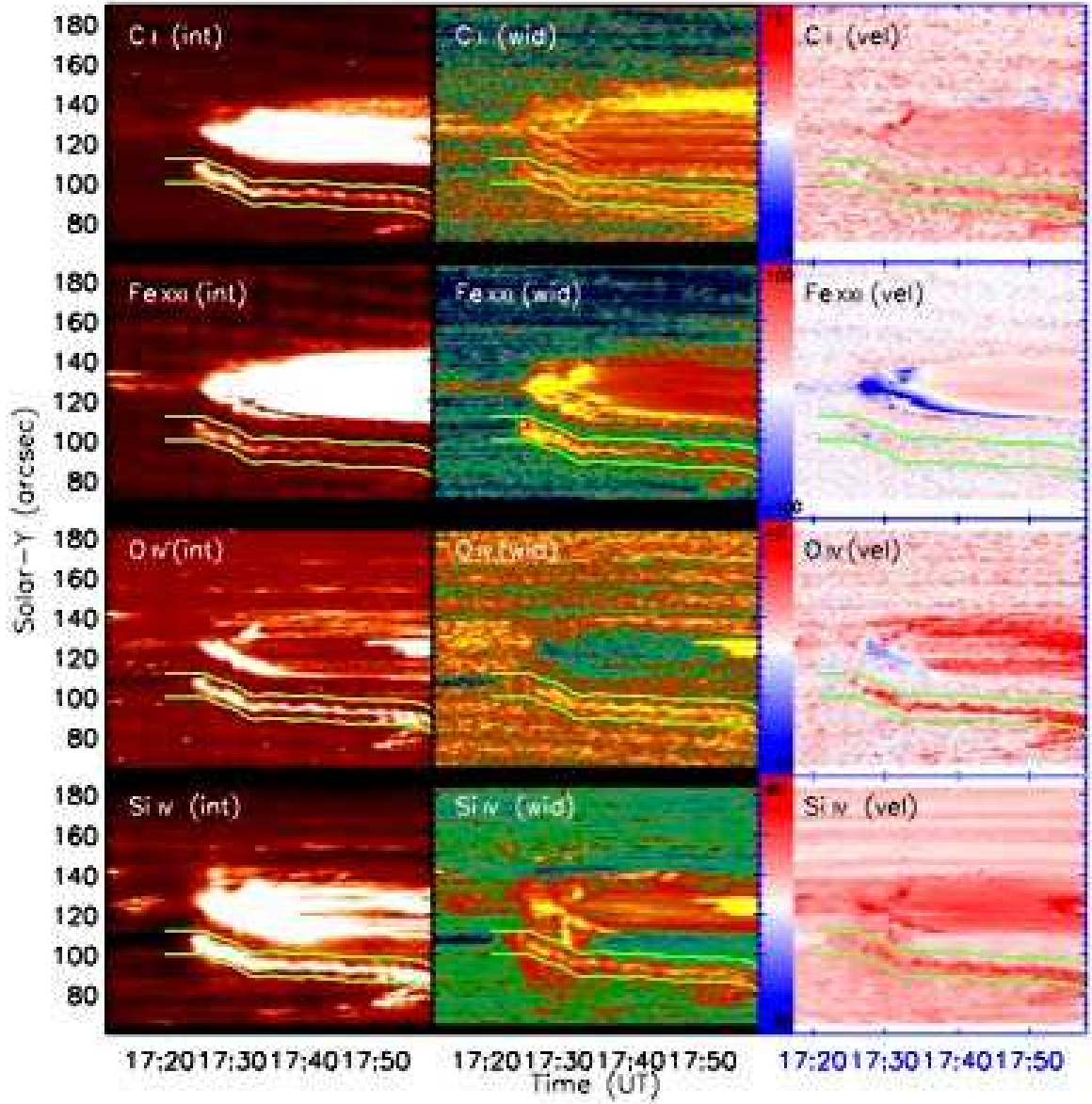} \caption{The space-time diagrams of
line intensity (left), line width (middle) and Doppler velocity
(right) from {\it IRIS} observations, such as \ci, \fexxi, \oiv, and
\siiv lines. The light curves integrated between two green lines are
shown in Fig.~\ref{qpp2}. \label{image1}}
\end{figure}

\begin{figure}
\epsscale{1.0} \plotone{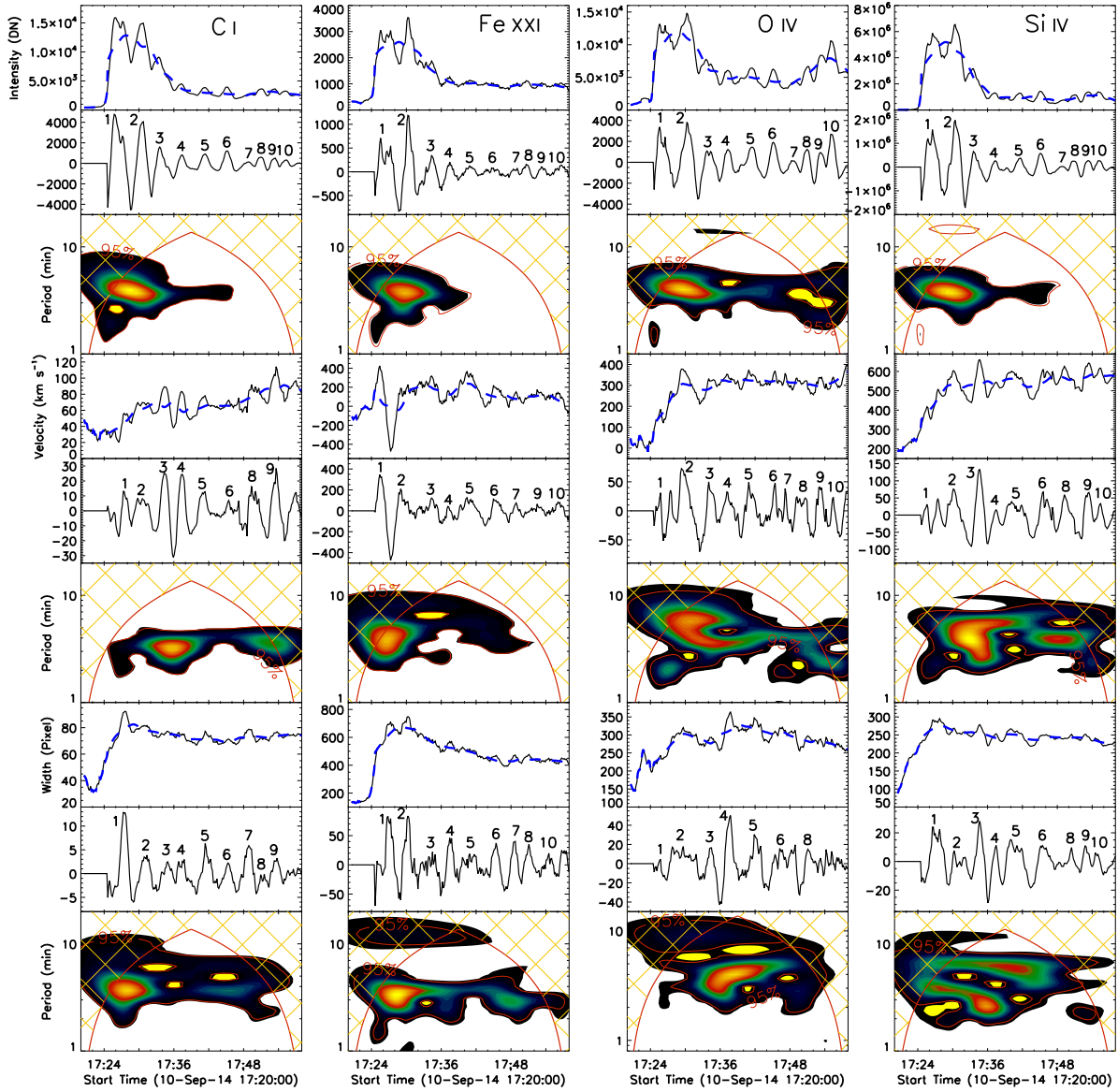} \caption{Time evolution of intensity
(row 1), Doppler velocity (row 4), and line width (row 7) at the
lines of \ci (column 1), \fexxi (column 2), \oiv (column 3), and
\siiv(column 4). The blue dashed lines represent the slowly-varying
components. Rows~(2, 5, 8): their rapidly-varying components.
Rows~(3, 6, 9): their wavelet power spectra. \label{qpp2}}
\end{figure}

Fig.~\ref{image1} shows the space-time diagrams of the line
intensity, line width and Doppler velocity after fitting four lines
at \ci 1354.33 {\AA}, \fexxi 1354.09 {\AA}, \oiv 1399.77 {\AA} and
\siiv 1402.77 {\AA}) from 17:12 to 17:58 UT. The Y-axis is along the
slit, which is fixed on the solar disk. This fact results into the
{\it IRIS} slit observes the same region of the flare ribbon during
this interval, as seen in the movie (f3.mpeg). There are two strong
emission patterns on these space-time diagrams. The northern one is
wide, and the southern pattern is narrow. This is because the slit
straddles on the curved flare ribbon, as shown in Fig.~\ref{active}.
These two emission patterns are just the different parts of this
ribbon. Different from the wide (northern) pattern, the narrow one
display the intensity variation with the time, including the \ci,
\fexxi, \oiv, and \siiv lines. This behavior is similar as the QPPs
shown in the HXR and radio emissions. In order to analyze this
feature in detail, we firstly use two lines to trace the southern
narrow emission pattern, as shown in Fig.~\ref{image1}. The distance
between these two lines is a constant of about 10$\arcsec$, as the
distance along the slit between two short green lines in
Fig.~\ref{active}. Secondly, the intensities between these two lines
are integrated. Thus, we get the intensity light curves of \ci,
\fexxi, \oiv, and \siiv lines. Fig.\ref{qpp2} shows that there are
ten peaks on their intensity light curves from 17:24 UT to 17:56 UT,
roughly with a quasi-period of less than 4 minutes. It is clear that
these peaks are superposed on a gradual component. Using the same
method as shown in Fig.~\ref{qpp1}, Fig.~\ref{qpp2} (upper-left
panel) shows that the \ci light curve is decomposed into a
slowly-varying and a rapidly-varying components. The slowly-varying
component is smoothing a window of 28 points. The rapidly-varying
component is the \ci light curve subtracting the slowly-varying
component. These ten peaks are clearly shown as marked by the
numbers, and they are identified as the QPPs. The wavelet spectra
confirm the quasi-period of less than 4 minutes. Such QPPs are also
distinctly detected from the \oiv and \siiv light curves. There are
some signatures of the QPPs on \fexxi light curves, especially the
first three peaks. The other seven peaks of \fexxi line are weak.
Fig.~\ref{qpp2} gives the temporal evolution of the Doppler
velocities and line widths at \ci, \fexxi, \oiv, and \siiv lines.
The flare ribbon displays the red shifts on the Doppler velocities
of these four lines. And the mean velocities between 17:24 UT and
17:58 UT are 67.4 km s$^{-1}$, 121.4 km s$^{-1}$, 297.9 km s$^{-1}$,
and 535.7 km s$^{-1}$ at \ci, \fexxi, \oiv, and \siiv lines, as
listed in table~\ref{tab}. There are also some peaks on the Doppler
velocities corresponding to that on the intensities, but not
one-by-one. The flare ribbon exhibits a broad line widths\, and the
mean values  between 17:24 UT and 17:58 UT are 73.9 pixels, 514.1
pixels, 291.2 pixels, and 247.3 pixels at \ci, \fexxi, \oiv, and
\siiv lines (see, table~\ref{tab}). There are some peaks on the line
widths corresponding to the QPPs peaks too. Using the same method,
their Doppler velocities and line widths are decomposed into the
slowly-varying and rapidly-varying components. The wavelet spectra
of the rapidly-varying components of Doppler velocity and line width
also exhibit the similar 4-min QPPs features.

\begin{figure}
\epsscale{1.0} \plotone{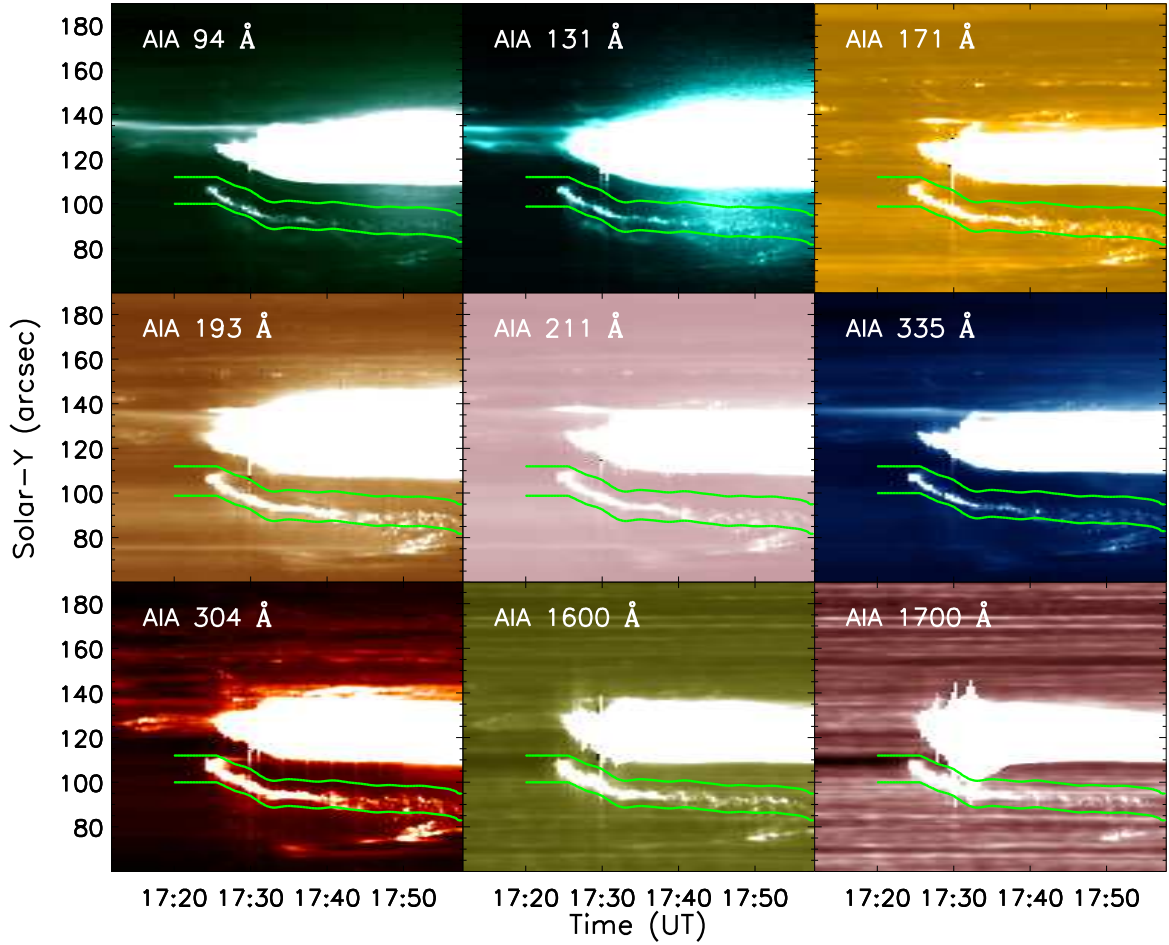} \caption{Same as Fig.~\ref{image1},
but the space-time diagrams at 9 AIA wavelengths from the same
position as the {\it IRIS} slit. \label{image2}}
\end{figure}

\begin{figure}
\epsscale{1.0} \plotone{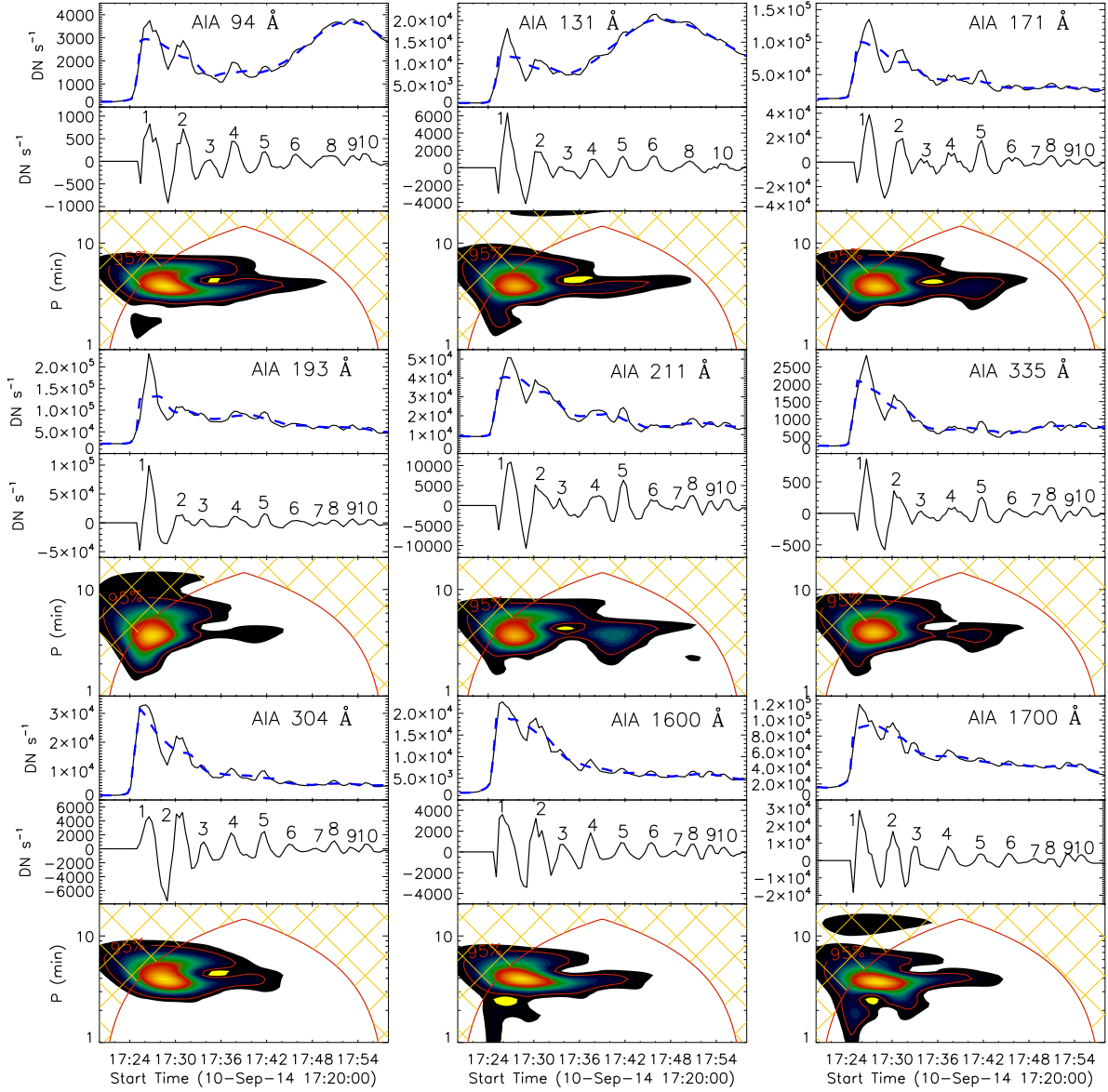} \caption{Rows~(1, 4, 7): the light
curves integrate between two green lines in Fig.~\ref{image2} at all
AIA 9 wavelengths. The blue dashed lines are the slowly-varying
components. Rows~(2, 5, 8): their rapidly-varying components.
Rows~(3, 6, 9): their wavelet power spectra. \label{qpp3}}
\end{figure}

{\it SDO}/AIA movie shows that the flare ribbon evolve from the
northeast toward the southwest on the solar disk, which results into
the flare ribbon cross the {\it IRIS} slits. The onset time at 17:24 UT in
Fig.~\ref{qpp2} is not the flare beginning time but represents the
starting time of the flare ribbon entrance into the slit window. It
is hard to detect the QPPs starting time from the flare ribbon.
However, an artificial slit with a position same as {\it IRIS} slit
is put on the AIA images. Thus we get the space-time diagrams from
AIA observations, as shown in Fig.~\ref{image2}. Same as the {\it
IRIS} observation in Fig.~\ref{image1}, there are two bright
patterns corresponding to the two parts of the flare ribbon. There
are intensity flashes at the southern small region. Using the same
two lines in Fig.~\ref{image1}, the light curves between them are
shown in Fig.~\ref{qpp3} at 9 AIA wavelengths. The QPPs are clearly
seen on their light curves. Ten individual peaks are recognized
between 17:24 UT and 17:56 UT, which is same as the {\it IRIS}
observations, and each peak of QPPs is marked with numbers. The
light curves are also decomposed into the slowly-varying (blue
dashed line) and rapidly-varying components, whose wavelet spectra
are also shown. The QPPs exhibit a period of less than 4 minutes
again, which is similar to that in the spectral lines at \ci, \fexxi,
\oiv, and \siiv from {\it IRIS} observations.

\section{Conclusions and Discussions}
Based on the multi-wavelengths observations from {\it Fermi}/GBM,
{\it SDO}/AIA, {\it IRIS} and SWAVES, we analyze the imaging and
spectral observations of the 4-min QPPs at HXR, EUV, and radio
emissions in a solar flare on 2014 September 10. We draw the
conclusions as following:

(I) The 4-min QPPs are found in a broad frequency range from HXR
through EUV to the radio.

(II) Imaging observations of {\it SDO}/AIA show that the QPPs
originate from the flare ribbon front.

(III) Spectral observations of {\it IRIS} present that the QPPs
peaks tend to a broad line width and a red-shift velocity.

Although there are several models to explain the QPPs in the
documents \citep[e.g.,][]{Aschwanden87,Nakariakov04,Nakariakov09},
our findings support that the QPPs are produced by the non-thermal
electron beams accelerated by the periodic magnetic reconnection in
this flare. In this case, each individual peak of the QPPs are
radiated by the different electron beams. Based on the standard
flare model, each magnetic reconnection can accelerate the
bi-directional electron beams simultaneously
\citep[e.g.,][]{Heyvaerts77,Aschwanden95,Innes97,Ning00,Ji06,Ji08,Shen08,Feng11,Feng13,Su13,Zhang12,Zhang14}.
The upward beam radiates the type III bursts on its trajectory
propagation outer the corona. The downward beam produces one peak at
HXR when it injects into the chromosphere to heat the local plasma
on its way producing one peak at EUV. In this case, the periodic
magnetic reconnection can produce the QPP peaks at HXR and radio
type III bursts, and the periodic EUV emissions as well. In other
words, the periodic magnetic reconnection model can well explain the
QPPs from HXR through EUV to the radio emissions in 2014 September
10 solar flare. In general, the other models also explain the QPPs
features, especially at HXR and EUV bands
\citep[e.g.,][]{Ofman02,Foullon05,Su12a}, i.e., the MHD flux tube
oscillations, modulated by certain waves or periodic self-organizing
systems of plasma instabilities. As mentioned earlier, solar type
III bursts are produced by the electron beams propagating into the
outer corona, then into the interplanetary. In other words, a group
of type III bursts are produced by various electron beams.
Therefore, the periodic solar type III bursts provide direct
evidence of the periodic magnetic reconnection in this flare.

\begin{figure}
\epsscale{1.0} \plotone{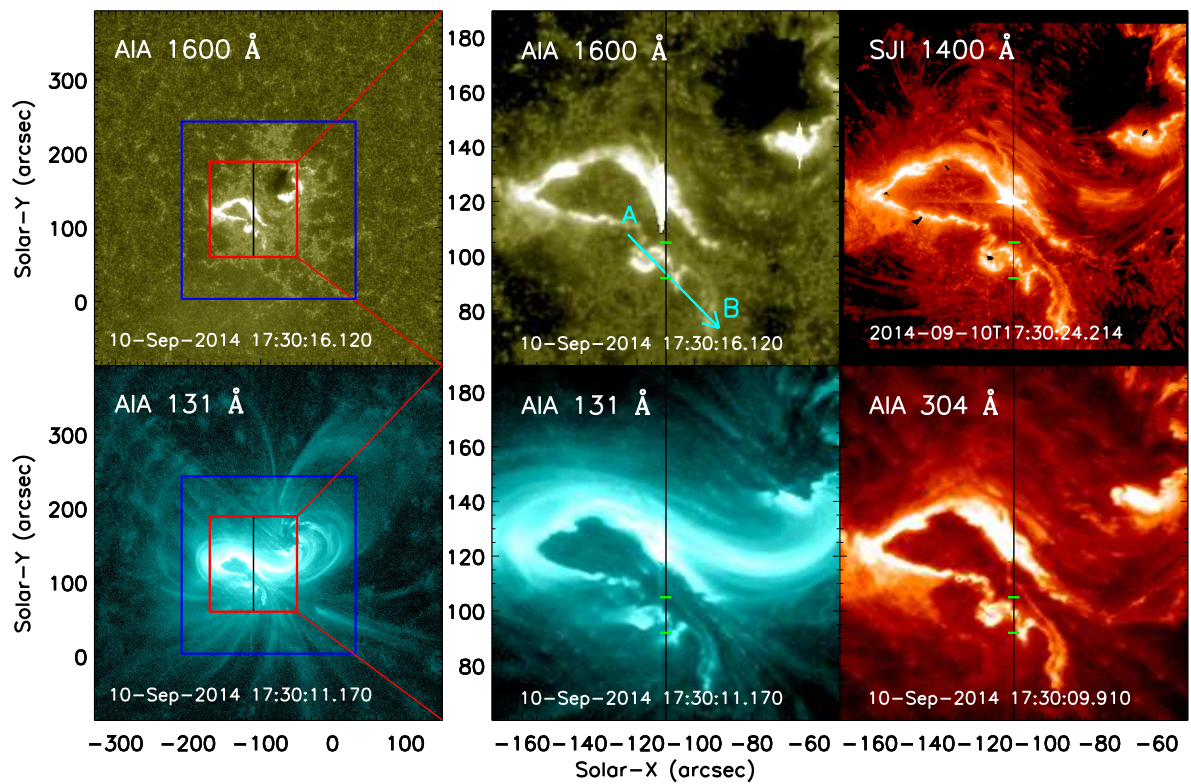} \caption{{\it SDO}/AIA and {\it
IRIS}/SJ images at 17:30 UT on 2014 September 10. Left: AIA images
at 1600 {\AA} and 131 {\AA}. The blue box marks the flare region to
integrate the light curves in Fig.~\ref{light}, and the red box give
the FOV of the SJ image. The vertical black lines mark the {\it
IRIS} slit positions. Right: AIA images at 1600 {\AA}, 131 {\AA},
and 304 {\AA} with the same FOV of SJ image at 1400 {\AA}. Two green
solid lines mark a distance of about 10$\arcsec$ along the slit (see
detail in the text). Arrow `A$\rightarrow$B' marks the slit of the
space-time diagram in Fig.~\ref{loop}. \label{active}}
\end{figure}

\begin{figure}
\epsscale{1.0} \plotone{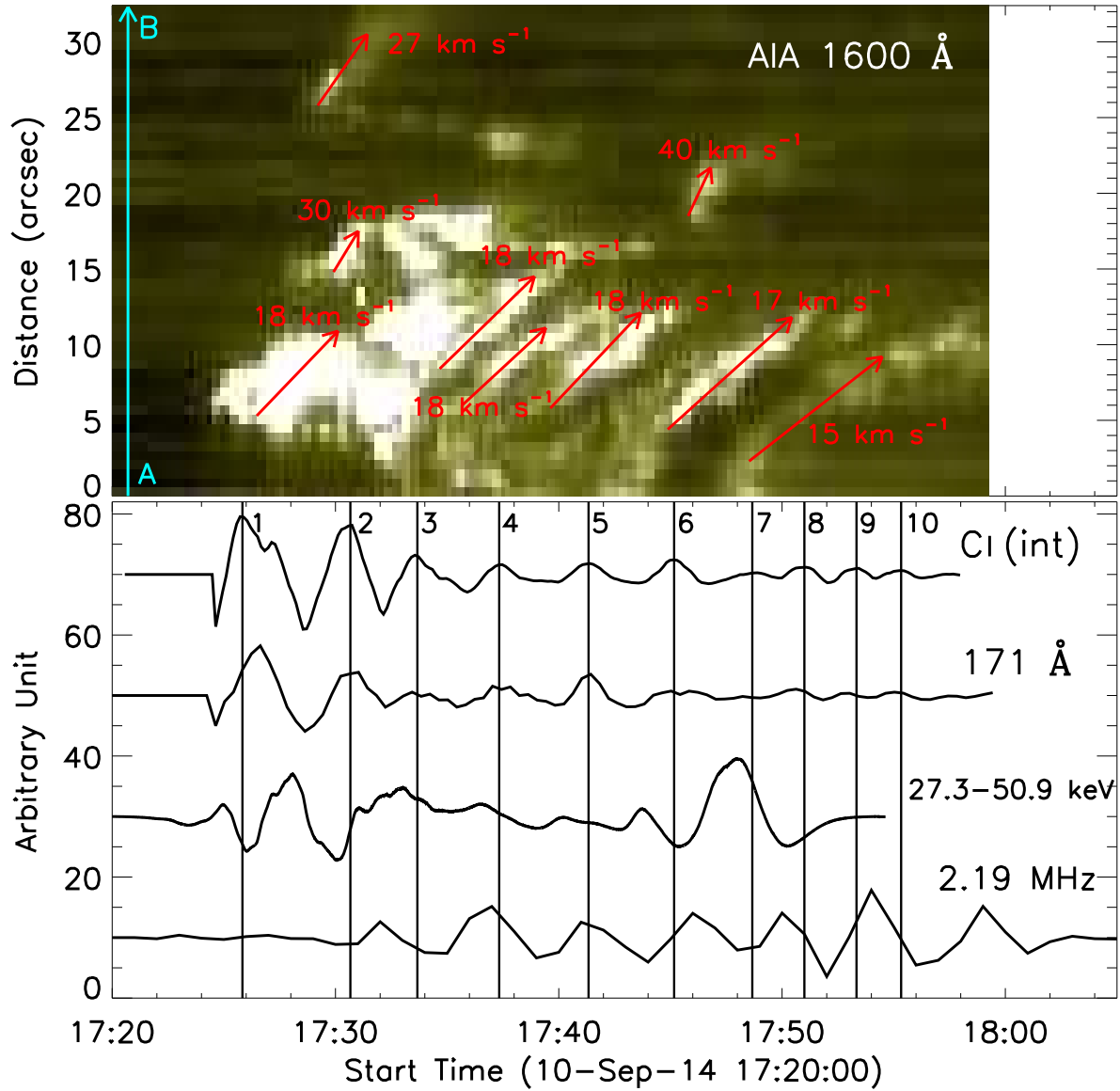} \caption{Upper: AIA 1600 {\AA}
space-time diagram along the slit `A$\rightarrow$B' in
Fig.~\ref{active}. Bottom: the illustration of the QPPs at {\it
IRIS} \ci line, AIA 171 {\AA}, {\it Fermi} 27.3$-$50.9 keV, and
SWAVES 2.19 MHz. \label{loop}}
\end{figure}

It is still an open question what does determine the period? The
quasi-periodic magnetic reconnection may be spontaneous
\citep{Kliem00,Karlicky04,Karlicky05,Murray09}, or could be
modulated by certain waves in solar corona
\citep{Aschwanden94s,Aschwanden04,Chen06,Ofman06,Nakariakov06,Inglis09,Nakariakov11,Liu11,Liting15}.
If in the case of spontaneous quasi-periodicity it is not understood
yet what determines the period \citep[e.g.,][]{Kliem00, Murray09}.
If in the latter induced case, the periodic triggering of the
reconnection can be some MHD oscillations, as were shown in several
papers \citep[e.g.,][]{Chen06,Nakariakov06,Nakariakov11}. The period
of the observational QPPs is thought to be associated with one of
the MHD modes. There are three possibilities for such MHD wave modes
which may trigger the periodic magnetic reconnection in our case.
The first possibility is the quasi-periodic reconnection modulated
by slow waves \citep{Chen06}. For example, the 3 or 5-min solar
p-mode oscillations (slow waves) can trigger the quasi-periodic
magnetic reconnection with the similar period when the reconnection
site is located in the upper chromosphere \citep{Ning04}. This is
similar to 4-min period of QPPs in the 2014 September 10 flare.
However, this model depends strongly on the location of the
reconnection site in the solar atmosphere, and it becomes weaker
when the reconnection site is lower or higher than the upper
chromosphere. The second possibility is the quasi-periodic
reconnection modulated by fast waves. \citet{Nakariakov06} suggest
that the fast waves in a corona loop situated near the flare site
can trigger the quasi-periodic magnetic reconnection. In particular,
the global kink mode \citep{Foullon05} can trigger the
quasi-periodic magnetic reconnection and produce the QPPs with a
period of about several minutes. This model is used to explain the
min-periodic phenomena in solar flares. The third case is the slow
magnetoacoustic waves to modulate the quasi-periodic reconnection.
\cite{Nakariakov11} have demonstrated that the slow magnetoacoustic
waves can propagate along the axis of a coronal magnetic arcade,
then possibly to trigger the quasi-periodic reconnection during the
solar flare. The observational period is similar to the period of a
standing slow magnetoacoustic wave in the loops that form the
arcade. They suggest that the quasi-periodic pulsations observed in
two-ribbon flares can be explained with this mechanism. Namely, such
mechanism can explain the 4-min period of QPPs in 2014 September 10
flare. This is because we find that the brightness structures move
along the flare ribbon parallel to the magnetic neutral line (seen
the movie of f3.mpeg). Fig.~\ref{loop} (upper panel) shows the
space-time slices along the flare ribbon A$\rightarrow$B on AIA 1600
\AA\ image in Fig.~\ref{active}. There are many brightening
structures moving from A to B. The propagation speed is roughly
estimated about 15$-$40 km s$^{-1}$, which is consistent with
previous findings
\citep[e.g.,][]{Bogachev05,Krucker05,Tripathi06,Li09,Reznikova10,Nakariakov11,Liting15}.
And these values are much smaller than the local Alfv\'{e}n and
sound speeds. The moving brightening structures are thought to be
the evidences of the slow magnetoacoustic waves across the magnetic
fields in solar flares \citep{Nakariakov11}.

The delay of about 8 minutes between the HXR (or EUV) and radio
emissions in this flare is possibly due to the different sites of
the radio sources (electron beams for type III bursts) from HXR or
EUV sources. Usually, HXR and EUV emissions are produced at the
chromosphere or lower corona, while the radio source around 2 MHz
originates from a heliocentric height of about 10R$_{\odot}$
\citep{Krupar14}. The time of the electron beams (type III bursts
source) propagating from the flare site (acceleration region) to
$\sim$10R$_{\odot}$ roughly equals to the delay between HXR (or EUV)
and radio emissions. Based on this assumption, we estimate that the
electron beams (radio sources) have a speed of $\sim$0.05 $c$ ($c$
is the light speed in the vacuum). This value is reasonable for the
electron beams propagate outward from the Sun in the interplanetary
medium \citep[e.g.,][]{Dulk87,Krupar14}. Fig.~\ref{loop} (bottom)
plots the QPPs at {\it IRIS} \ci intensity, AIA 171 {\AA}, {\it
Fermi} 27.3$-$50.3 keV, and SWAVES 2.19 MHz. Each peaks of the QPPs
at \ci intensity is well corresponding to that at AIA 171 {\AA}.
However, they are not correlated to the HXR and radio peaks, which
could be resulted from the radiation source positions.

\acknowledgments \textbf{The authors would like to thank the
anonymous referee for his/her valuable comments to improve the
manuscript. The data used in this paper are from {\it Fermi}, {\it
STEREO}, {\it SDO}/AIA, {\it IRIS}, and {\it GOES}. This study is
supported by NSFC under grants 11073058, 11073006, 11173062,
11203083, 11333009, 11303101, 11473071, 973 program (2011CB811400,
2014CB744200) and Laboratory NO. 2010DP173032. We are grateful to
Dr. D.E.~Innes for {\it IRIS} data, Drs. I.~Sharykin, A.~Struminsky,
and W.~Chen for {\it Fermi} data, Dr. Y.~Huang for {\it STEREO}
data, Dr. L.~Xu for helpful discussions.}


\begin{table}
\caption{The mean value and standard deviation on Doppler velocities
and line widths between 17:24 UT and 17:58 UT of the spectral lines
from {\it IRIS} observations.} \centering
\begin{tabular}{c c c c c}
 \hline\hline
Spectrum  & \multicolumn{2}{c}{Doppler velocity (km s$^{-1}$)}  & \multicolumn{2}{c}{Line width (pixels)} \\
          &  Mean  & Standard deviation            &  Mean & Standard deviation  \\
  \hline
\ci       & 67.4   &  17.9   &  73.9   & 5.1   \\
\fexxi    & 121.4  &  140.4  &  514.1  & 91.2  \\
\oiv      & 297.9  &  55.7   &  291.2  & 27.8  \\
\siiv     & 535.7  &  65.9   &  247.3  & 16.8  \\
\hline
\end{tabular}
\label{tab}
\end{table}

\clearpage
\end{document}